\documentclass[runningheads]{llncs}
\usepackage{multirow}
\usepackage{subcaption}
\usepackage{amssymb}
\usepackage{booktabs}
\usepackage{graphicx}
\usepackage[misc]{ifsym}
\usepackage[T1]{fontenc}
\usepackage{graphicx}
\usepackage{hyperref}
\usepackage{color}

\usepackage[final]{pdfpages}    
\usepackage{tabularx}

\begin{document}
\title{ASPS: Augmented Segment Anything Model for Polyp Segmentation}
%
%
\author{
Huiqian Li\inst{1,2} \and   
Dingwen Zhang\inst{2,3} $^{(\textrm{\Letter})}$ \and    
Jieru Yao\inst{2,3} \and        
Longfei Han\inst{4, 1} $^{(\textrm{\Letter})}$ \and      
Zhongyu Li\inst{5} \and         
Junwei Han\inst{2,3} } 
\authorrunning{H. Author et al.}
%
\institute{University of Science and Technology of China, Hefei, China \\ \and
Institute of Artificial Intelligence, Hefei Comprehensive National Science Center, Hefei, China\\ \and
Northwestern Polytechnical University, Xi'an, China \\ \and
Beijing Technology and Business University, Beijing, China \\ \and
Xi'an Jiaotong University, Xi'an, China \\
\email{zhangdingwen2006yyy@gmail.com, draflyhan@gmail.com}
}
%
\maketitle              
\begin{abstract}
Polyp segmentation plays a pivotal role in colorectal cancer diagnosis. Recently, the emergence of the Segment Anything Model (SAM) has introduced unprecedented potential for polyp segmentation, leveraging its powerful pre-training capability on large-scale datasets. However, due to the domain gap between natural and endoscopy images, SAM encounters two limitations in achieving effective performance in polyp segmentation. Firstly, its Transformer-based structure prioritizes global and low-frequency information, potentially overlooking local details, and introducing bias into the learned features. Secondly, when applied to endoscopy images, its poor out-of-distribution (OOD) performance results in substandard predictions and biased confidence output. 
To tackle these challenges, we introduce a novel approach named \textbf{A}ugmented \textbf{S}AM for \textbf{P}olyp \textbf{S}egmentation (\textbf{ASPS}), equipped with two modules: Cross-branch Feature Augmentation (CFA) and Uncertainty-guided Prediction Regularization (UPR). CFA integrates a trainable CNN encoder branch with a frozen ViT encoder, enabling the integration of domain-specific knowledge while enhancing local features and high-frequency details. Moreover, UPR ingeniously leverages SAM's IoU score to mitigate uncertainty during the training procedure, thereby improving OOD performance and domain generalization. 
Extensive experimental results demonstrate the effectiveness and utility of the proposed method in improving SAM's performance in polyp segmentation. Our code is available at \href{https://github.com/HuiqianLi/ASPS}{https://github.com/HuiqianLi/ASPS}. 

\keywords{Polyp Segmentation  \and Segment Anything Model \and Domain Adaptation.}
\end{abstract}
\section{Introduction}
Automated polyp segmentation stands as a pivotal tool in the diagnosis of colorectal cancer, to aid effective interventions and timely treatment strategies. 
Studies like Polyp-PVT\cite{dong2021polyp}, SSFormer\cite{wang2022stepwise} used Pyramid Vision Transformer for polyp segmentation; CFANet\cite{zhou2023cross} integrated boundaries with a Cross-level Feature Aggregation Network; Endo-FM\cite{wang2023foundation} captured spatial-temporal dependencies to build a foundation model.
However, limited by the model's size, the existing methods still lack sufficient capabilities for feature representation and extraction, making it challenging to fully capture the morphology and characteristics of polyps. Furthermore, the limited scale of the dataset may limit the diversity and generalization of the existed methods.
Recently, the Segment Anything Model\cite{kirillov2023segment} (SAM) was introduced. SAM is pre-trained on the largest segmentation dataset SA-1B, demonstrating remarkable segmentation capabilities across various downstream tasks. With its significant model size and data size, this innovative approach has introduced novel perspectives to the field of polyp segmentation. It also possesses enhanced representation and feature extraction capabilities, surpassing existing methods. 

However, SAM's performance in the polyp segmentation task is unsatisfactory \cite{zhou2023can}, due to the domain gap between the training data and endoscopy images. This results in two primary issues: firstly, SAM fails to adequately capture the distinctive features of polyp images, leading to a bias in its learned representations. Secondly, it produces erroneous predictions with inaccurate confidence estimates for out-of-distribution (OOD) data. In addition, because it relied on prompts, SAM has significantly impeded its convenience in clinical applications. Despite several methods improving SAM, such as MedSAM\cite{ma2024segment}, these approaches either rely on prompts or directly fine-tune substantial models. SAMUS\cite{lin2023samus} effectively integrates CNN and ViT, but its design is quite complex and is particularly suited for processing small images. Consequently, the efficacy of these methods is somewhat constrained. Various methods have been proposed to tackle the challenge of unsupervised domain adaptation in semantic segmentation. MIC\cite{hoyer2023mic} proposed a Masked Image Consistency module for target domain context learning; Context-Aware Domain Adaptation\cite{yang2021context} improved context transfer via cross-attention. Yet, domain-specific information integration and uncertainty reduction are still unexplored.

To address these issues, we introduce a novel method based on SAM from a domain adaptation perspective, designed to augment the feature extraction capability and generalization without relying on prompts. We propose the Cross-branch Feature Augmentation Module (CFA) and the Uncertainty-guided Prediction Regularization Module (UPR). CFA incorporates an additional trainable convolutional neural network (CNN) encoder branch, which complements the frozen vision transformer (ViT) encoder, to capture multi-scale and multi-level features. UPR adjusts the normalization layer to promote the adaptation in the endoscopy field and leverages hints to ensure accurate confidence estimation, so as to improve the OOD performance of SAM. 

In summary, our primary contributions are as follows:
(1) We build a novel SAM-based model named ASPS, to enhance the feature learning capability and domain generalization for polyp segmentation, demonstrating strong performance without the need for prompts.
(2) We introduce the Cross-branch Feature Augmentation Module (CFA), which introduces an additional CNN encoder branch as a supplement to the ViT encoder. Furthermore, we propose the Uncertainty-guided Prediction Regularization Module (UPR), leveraging hints to reduce uncertainty during training and improve the domain generalization of SAM.
(3) Extensive experiments on five common polyp datasets demonstrate the effectiveness and superiority of our method.

\section{Method}
\begin{figure}[t]
\includegraphics[width=\textwidth]{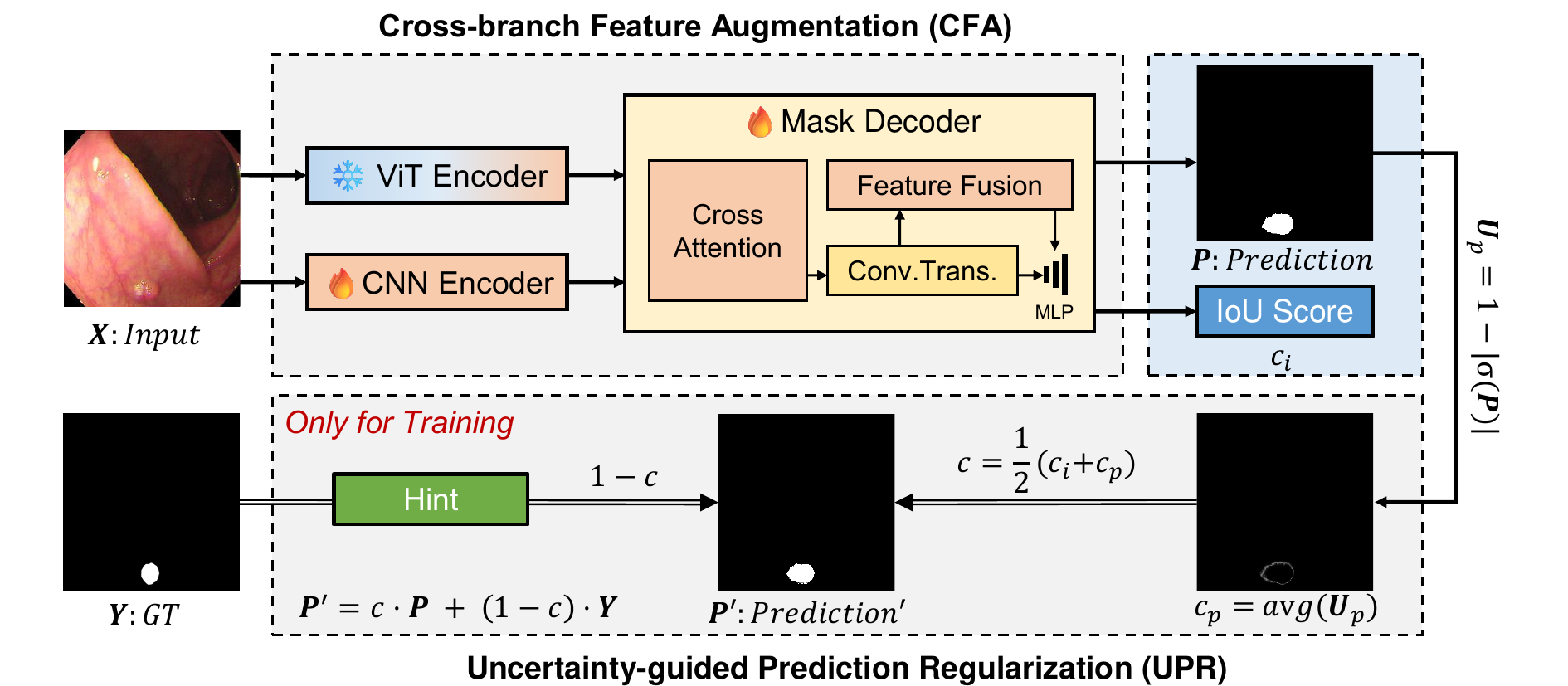}
\caption{An overview of our Augmented Segment Anything Model for polyp segmentation. The Cross-branch Feature Augmentation module is encouraged to learn multi-scale features and multi-level representations. The Uncertainty-guided Prediction Regularization module is designed to minimize the uncertainty of the prediction to improve the domain generalization ability of the model.} \label{framework}
\end{figure}

\textbf{Overview.}
Our proposed network is illustrated in Fig.~\ref{framework}. To address the domain degradation issue of SAM, we leverage two modules to enhance its original feature extraction capabilities and domain generalization. The CFA module integrates the CNN encoder feature with global ViT information, leading to generalized feature representation learning. This integration facilitates refined segmentation outputs by aggregating deep information to the superficial layers and incorporating positional information from the shallow levels. Meanwhile, the UPR module is designed to minimize uncertainty and calibrate confidence during training. UPR utilizes a training strategy based on uncertainty, leveraging the ground truth as a guiding `hint'. The proposed network follows end-to-end training without prompts, jointly optimizing two modules to achieve optimal performance.


\subsection{Cross-branch Feature Augmentation Module}
While SAM has achieved great success in many image segmentation tasks, it still has some limitations in the polyp segmentation task. One of the main reasons is that the image encoder of SAM is not able to capture enough features effectively from unseen endoscopy images. To address this issue, the CFA module is designed to learn multi-scale features and multi-level representations, thereby enhancing the feature extraction capabilities of the encoder. 

\begin{figure}[t]
\includegraphics[width=\textwidth]{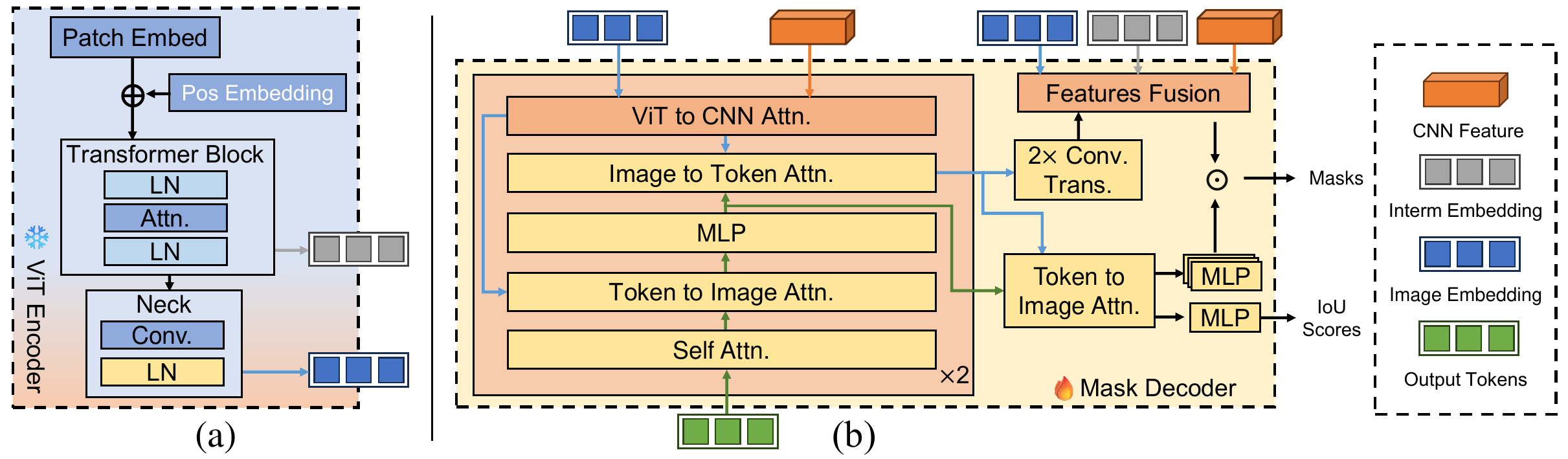}
\caption{Detailed architecture of ViT Encoder and Mask Decoder. (a) represents the ViT encoder, while (b) showcases the lightweight decoder of SAM. The CNN feature is derived from the output of the CNN encoder. The yellow and blue modules represent the original SAM structure.} \label{decoder}
\end{figure}

Firstly, to achieve automatic segmentation, we modified the architecture of SAM by removing its prompt input and prompt encoder components while preserving its image encoder and mask decoder parts. Recent studies\cite{pan2022fast} have demonstrated that ViT is more focused on low-frequency signals, while CNN is more adept at processing high-frequency signals. Hence, we integrate a parallel CNN-based branch to compensate for the absence of high-frequency and local features. Furthermore, we augment the mask decoder of SAM by proposing an additional multi-head cross-branch attention block to facilitate the integration of features extracted from both the ViT encoder and the CNN encoder. For the features ${\bf{F}}_v$ from the ViT branch and ${\bf{F}}_c$ from the CNN, the cross-branch attention can be formulated as follows: 
\begin{equation}
\textrm{CrossBranchAttention}\left({\bf{F}}_v,{\bf{F}}_c \right) = \textrm{Softmax}\left(\frac{{\bf{QK}}^{\mathsf{T}}}{\sqrt{d}}\right)\bf{V} .\label{eq1} 
\end{equation}
where $\bf{Q}={\bf{F}}_\mathit{v} \bf{W^Q}$, $\bf{K}=\bf{V}={\bf{F}}_\mathit{c} \bf{W^K}$, and $d$ is the number of channels of each head of ${\bf{F}}_v$. Considering that CNN features offer more precise position information, we substituted SAM's original position embedding in the mask decoder with the final output features from the CNN encoder. Furthermore, we integrated the cross-branch attention mechanism into the attention block of the mask decoder, repeating this process twice to ensure integration of multi-scale features from both the ViT and CNN encoders, as shown in Fig.~\ref{decoder}.

Secondly, to obtain more precise segmentation results, we integrate high-level context and low-level boundary information from the encoder with the decoder features of SAM to augment the output information. Specifically, we combine the shallow local features obtained from the intermediate embedding of the ViT encoder, the final global features obtained from the image embedding of the ViT encoder, and the final features of the CNN encoder, as illustrated in Fig.~\ref{decoder}. This approach fully harnesses the rich edge information, extensive global context information, and local position details of each encoder branch. Thus, we can effectively integrate multi-level features from both ViT and CNN. 

\subsection{Uncertainty-guided Prediction Regularization Module}
To augment the generalization capability of SAM, we propose a novel training strategy involving the selective activation of LayerNorm within the encoder. We also employ the ground truth as a `hint' to further guide the training process by correcting the confidence.

Given that SAM is trained on natural data, its performance may deteriorate in polyp images due to the domain transfer. As previously suggested \cite{li2016revisiting}, it is a particularly effective technique for domain transfer by adjusting the normalization layer. 
Despite the introduction of LayerNorm \cite{ba2016layer} potentially reducing training time, it fundamentally alters the distribution of the input data. When transferring SAM from natural images to endoscopy images, there is a shift in both the data distribution and the corresponding feature space distribution. These distributional differences can induce internal covariate shifts, thereby influencing the model's performance. To improve the SAM's generalization in the endoscopy field, we fine-tune the normalization layer of the encoder. In this process, the model effectively adapts the data distribution in the target domain and mitigates the effects of internal covariate shifts.

Specifically, the LayerNorm of SAM's ViT encoder is divided into (1) Transformer block norm, and (2) neck layer norm, as illustrated in Fig.~\ref{decoder}(a). Given that the features in the neck layer are closer to the output features of the encoder, we ultimately decided to train the neck layer normalization, which is equivalent to re-normalizing the features of the pre-trained ViT encoder. Coincidentally, in this work\cite{zhao2023tuning}, the straightforward technique of adjusting normalization layers can surprisingly yield comparable or even superior performance to the robust baseline of fine-tuning all parameters.

Moreover, previous research\cite{nguyen2023out} has demonstrated that predictions with lower uncertainty tend to exhibit superior out-of-distribution (OOD) performance, which is also beneficial for domain adaptation. SAM generates an IoU score output, which inherently can represent uncertainty (or, conversely, confidence). However, during the prediction process, SAM may frequently produce incorrect predictions for unseen data with high confidence, which is undesirable. To mitigate this problem, we strive to reduce the uncertainty of the model during training (i.e., to increase confidence). Inspired by \cite{devries2018learning}, we utilize the ground truth as a hint to guide the learning of the model. First, we represent the IoU score of SAM as the image-level confidence $c_i$. Then we calculate the pixel-level confidence $c_p$ to refine the uncertainty of each pixel using Eq.~\ref{eq_cp}, where ${\bf{U}}_p \in \mathbb{R}^{B\times 1 \times H\times W}$. 
\begin{equation}
c_p = \left( 1-\frac{1}{H\times W} \sum_{i}^{H}  \sum_{j}^{W} {\bf{U}}_p \right)  .\label{eq_cp}
\end{equation}

The term ${\bf{U}}_p$ represents the pixel uncertainty, defined as ${\bf{U}}_p = 1 - \sigma(\left| \bf{P} \right|)$. Here, $\sigma$ represents the Sigmoid function, while $\bf{P}$ represents the output prediction. The final confidence is calculated as the sum of the image-level confidence and the pixel-level confidence, expressed as $c=\frac{1}{2} \left(c_i+c_p\right) $. This confidence is determined by the Bernoulli distribution, which decides whether to utilize the ground truth as a hint. In other words, if the confidence is low enough, we believe that the model requires a specific answer hint to learn the correct mask prediction. Thus, the answer is required as a hint, otherwise, it is not necessary. The weight of the hint is determined by the confidence $c$, which is expressed as follows: 
\begin{equation}
\bf{P'} = \mathit{c} \cdot \bf{P} + (\textrm{1}-\mathit{c}) \cdot \bf{Y} .\label{eq3}
\end{equation}

However, by minimizing the loss function, the model will tend to make $c=0$ so that $\bf{P'}$ will always be GT. This means that the model does not actually learn. Therefore, a confidence loss is introduced to supervise $c$, which will increase when $c\to 0$, and the confidence loss is defined as follows: 
\begin{equation}
\mathcal{L}_c = - log(c) .\label{eq4}
\end{equation}

The final loss function is the sum of the segmentation loss $\mathcal{L}_s$ and the confidence loss $\mathcal{L}_c$, as defined in Eq.~\ref{eq5}. Here, $\lambda$ represents a hyperparameter. Specifically, the segmentation loss employed is a combination of CE loss, Dice loss, and MSE loss as $\mathcal{L}_s = \mathcal{L}_{ce} + 0.5 \cdot \mathcal{L}_{dice} +  \mathcal{L}_{mse}$. 
\begin{equation}
\mathcal{L} = \mathcal{L}_s + \lambda \mathcal{L}_c .\label{eq5}
\end{equation}
 
\section{Experiments}
\subsubsection{Datasets.}
We conduct extensive experiments on five polyp segmentation datasets following PraNet\cite{fan2020pranet}, including Kvasir-SEG\cite{jha2020kvasir}, CVC-ClinicDB\cite{bernal2015wm}, CVC-ColonDB\cite{tajbakhsh2015automated}, ETIS\cite{silva2014toward} and EndoScene\cite{vazquez2017benchmark}. Specifically, the training set consists of 900 images from Kvasir-SEG and 550 images from ClinicDB. The test sets comprise 100 images from Kvasir-SEG, 62 images from CVC-ClincDB, 380 images from CVC-ColonDB, 60 images from EndoScene, and 196 images from ETIS.

\subsubsection{Implementations. }
We use PyTorch to implement our method and conduct experiments on a single NVIDIA RTX3090 GPU. The AdamW optimizer is utilized for training 16K iterations with a learning rate of 1e-5, a weight decay of 1e-4, and a batch size of 4. The CNN model we utilize is MSCAN-L, sourced from SegNeXt\cite{guo2022segnext}. The input image size for the ViT branch is $1024\times1024$, while the input size for the CNN branch is $320\times320$. In the evaluation stage, we use two common metrics in medical image segmentation, \textit{Dice} and \textit{IoU}. 

\subsubsection{Results and Analysis. }
We compare our method with several state-of-the-art polyp segmentation methods and some SAM-based methods in Table~\ref{tab1}. It is evident that while the performance enhancement on Kvasir and CVC-ColonDB is not particularly noticeable, the improvement on CVC-ClinicDB, ETIS, and EndoScene is quite significant. Especially, Polyp-PVT\cite{dong2021polyp} showed good performance on all datasets with average Dice and IoU of 0.870 and 0.804, respectively, while our method achieved 0.890 and 0.817, which proves the effectiveness of our model. 

Compared to SAM-based (ViT-B) methods, our approach outperformed all others across all datasets. It's important to note that methods like MedSAM\cite{ma2024segment} and SAMUS\cite{lin2023samus} still incorporated prompts like SAM, but we removed prompts in the comparative experiment. Besides, despite our efforts, we didn't attain satisfactory results with SurgicalSAM\cite{yue2023surgicalsam} and SAMed\cite{zhang2023customized}, potentially due to incompatible training hyperparameters. Moreover, we observed that SAMUS\cite{lin2023samus} also utilized a CNN auxiliary branch and achieved excellent results, respectively, further validating the effectiveness of the CNN branch. Additionally, with the development of the lightweight model of SAM, we successfully combined our method with EfficientSAM\cite{xiong2023efficientsam} and achieved satisfactory results.

\begin{table}[t]
\caption{Quantitative comparisons with state-of-the-art (SOTA) methods on five public polyps datasets are presented. We mark the best results with \textbf{bold} and the second-best scores with \underline{underline}. }\label{tab1}
\resizebox{\linewidth}{!}{  
\begin{tabular}{cccccccccccc}
\toprule
\multirow{2}{*}{\textbf{Methods}} & \multicolumn{1}{c|}{\multirow{2}{*}{\textbf{\begin{tabular}[c]{@{}c@{}}Published\\ Venue\end{tabular}}}} & \multicolumn{2}{c}{\textbf{\makebox[2.25cm]{CVC-ClinicDB}}} & \multicolumn{2}{c}{\textbf{\makebox[2cm]{Kvasir}}} & \multicolumn{2}{c}{\textbf{\makebox[2cm]{CVC-ColonDB}}} & \multicolumn{2}{c}{\textbf{\makebox[2cm]{ETIS}}} & \multicolumn{2}{c}{\textbf{\makebox[2cm]{EndoScene}}} \\
& \multicolumn{1}{c|}{} & Dice & IoU & Dice & IoU & Dice & IoU & Dice & IoU & Dice & IoU \\
\midrule
UNet\cite{ronneberger2015u} & \multicolumn{1}{c|}{MICCAI'15} & 0.823 & 0.755 & 0.818 & 0.746 & 0.504 & 0.436 & 0.398 & 0.335 & 0.710 & 0.627 \\
PraNet\cite{fan2020pranet} & \multicolumn{1}{c|}{MICCAI'19} & 0.899 & 0.849 & 0.898 & 0.840 & 0.709 & 0.640 & 0.628 & 0.567 & 0.871 & 0.797 \\
SANet\cite{wei2021shallow} & \multicolumn{1}{c|}{MICCAI'21} & 0.916 & 0.859 & 0.904 & 0.847 & 0.752 & 0.669 & 0.750 & 0.654 & 0.888 & 0.815 \\
MSNet\cite{zhao2021automatic} & \multicolumn{1}{c|}{MICCAI'21} & 0.915 & 0.866 & 0.902 & 0.847 & 0.747 & 0.668 & 0.720 & 0.650 & 0.862 & 0.796 \\
UACANet\cite{kim2021uacanet} & \multicolumn{1}{c|}{MM'21} & 0.916 & 0.870 & 0.905 & 0.852 & 0.783 & 0.704 & 0.694 & 0.615 & 0.902 & 0.837 \\
LDNet\cite{zhang2022lesion} & \multicolumn{1}{c|}{MICCAI'22} & 0.932 & 0.872 & 0.912 & 0.855 & 0.794 & 0.715 & 0.778 & 0.707 & 0.893 & 0.826 \\
SSFormer\cite{wang2022stepwise} & \multicolumn{1}{c|}{MICCAI'22} & 0.906 & 0.855 & \underline{0.917} & \textbf{0.864} & \underline{0.802} & \underline{0.721} & 0.796 & 0.720 & 0.895 & 0.827 \\
DCRNet\cite{yin2022duplex} & \multicolumn{1}{c|}{ISBI'22} & 0.869 & 0.800 & 0.846 & 0.772 & 0.661 & 0.576 & 0.509 & 0.432 & 0.753 & 0.670 \\
Polyp-PVT\cite{dong2021polyp} & \multicolumn{1}{c|}{AIR’23} & 0.937 & 0.889 & \underline{0.917} & \textbf{0.864} & \textbf{0.808} & \textbf{0.727} & 0.787 & 0.706 & 0.900 & 0.833 \\
CFANet\cite{zhou2023cross} & \multicolumn{1}{c|}{PR'23} & 0.933 & 0.883 & 0.915 & \underline{0.861} & 0.743 & 0.665 & 0.732 & 0.655 & 0.893 & 0.827 \\
\midrule
SAM-H\cite{kirillov2023segment} & \multicolumn{1}{c|}{ICCV'23} & 0.547 & 0.500 & 0.778 & 0.707 & 0.441 & 0.396 & 0.517 & 0.477 & 0.651 & 0.606 \\
SAM-L\cite{kirillov2023segment} & \multicolumn{1}{c|}{ICCV'23} & 0.579 & 0.526 & 0.782 & 0.710 & 0.468 & 0.422 & 0.551  & 0.507 & 0.726 & 0.676   \\
SAM-Adapter\cite{chen2023sam} &  \multicolumn{1}{c|}{ICCV'23}    & 0.774 & 0.673    & 0.847 & 0.763 & 0.671  & 0.568  & 0.590 & 0.476  & 0.815 & 0.725 \\
AutoSAM\cite{hu2023efficiently}  &\multicolumn{1}{c|}{ArXiv'23} & 0.751& 0.642 & 0.784 & 0.675 & 0.535  & 0.418 & 0.402& 0.308 & 0.829 & 0.739 \\
SAMPath\cite{zhang2023sam} &\multicolumn{1}{c|}{MICCAIw'23} & 0.750 & 0.644& 0.828 & 0.730 & 0.632 & 0.516 & 0.555 & 0.442 & 0.844 & 0.756 \\
SAMed\cite{zhang2023customized} & \multicolumn{1}{c|}{ArXiv'23} & 0.404 & 0.273 & 0.459 & 0.300 & 0.199 & 0.115 & 0.212 & 0.126 & 0.332 & 0.202 \\
SAMUS\cite{lin2023samus} & \multicolumn{1}{c|}{ArXiv'23} & 0.900 & 0.821 & 0.859 & 0.763 & 0.731 & 0.597 & 0.750 & 0.618 & 0.859 & 0.760 \\
SurgicalSAM\cite{yue2023surgicalsam}  & \multicolumn{1}{c|}{AAAI'24} & 0.644& 0.505& 0.740 & 0.597 & 0.460& 0.330 & 0.342 & 0.238 & 0.623  & 0.472  \\
MedSAM\cite{ma2024segment} & \multicolumn{1}{c|}{Nature'24} & 0.867 & 0.803 & 0.862 & 0.795 & 0.734 & 0.651 & 0.687 & 0.604 & 0.870 & 0.798 \\
\midrule
Ours & \multicolumn{1}{c|}{Efficient-SAM\cite{xiong2023efficientsam}} &0.942 & 0.891 & 0.914 & 0.849 & 0.782 & 0.680 & 0.854 & 0.758 & 0.900 & 0.819\\
Ours & \multicolumn{1}{c|}{ViT-B} & \underline{0.950} & \underline{0.905} & 0.914 & 0.848 & 0.792 & 0.694 & \underline{0.856} & \underline{0.764} & \underline{0.914} & \underline{0.843} \\
Ours & \multicolumn{1}{c|}{ViT-H} & \textbf{0.951} & \textbf{0.906} & \textbf{0.920} & 0.858 & 0.799 & 0.701 & \textbf{0.861} & \textbf{0.769} & \textbf{0.919} & \textbf{0.852} \\
\bottomrule
\end{tabular}
}
\end{table}

\begin{figure}[t]
\centering
\begin{minipage}[t]{0.45\textwidth}
\centering
\includegraphics[width=\textwidth]{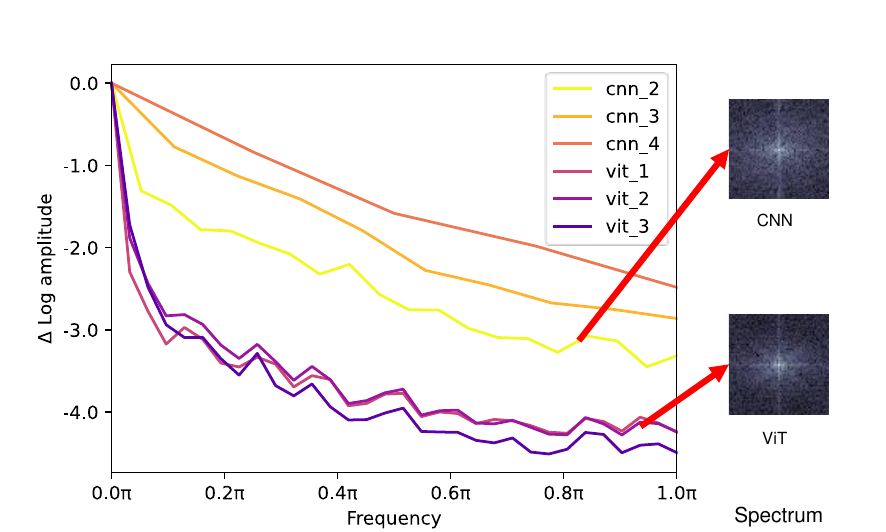}
\caption{Relative log amplitudes.} \label{fourier}
\end{minipage}
\begin{minipage}[t]{0.50\textwidth}
\centering
\includegraphics[width=\textwidth]{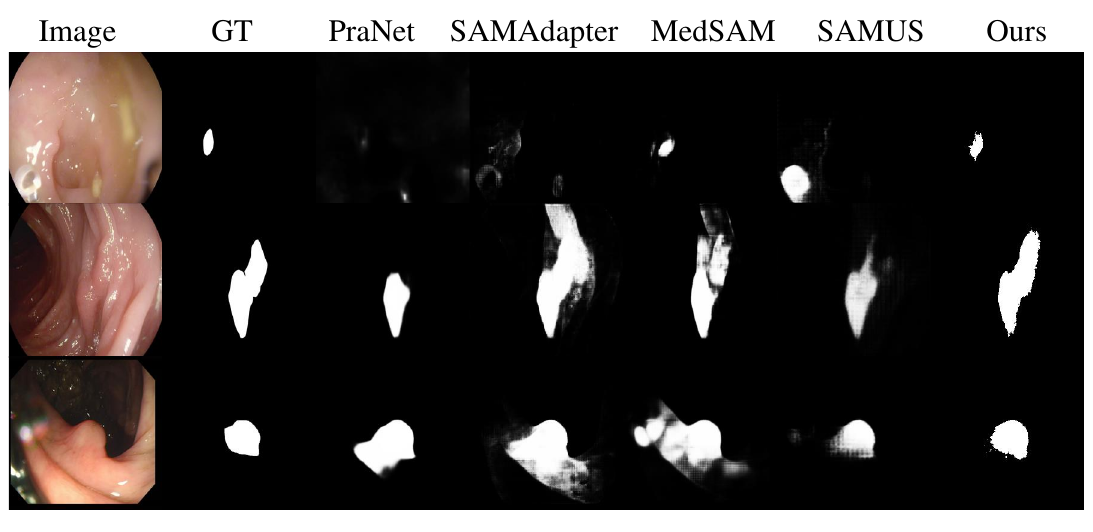}
\caption{Several qualitative results.} \label{qualitative}
\end{minipage}
\end{figure}

In Fig.~\ref{fourier}, we use Fourier analysis as a toolkit to show the difference between features from two encoders. The Fourier spectrum and relative log amplitudes of the Fourier transformed feature maps indicate that the CNN branch captures more high-frequency signals than the ViT baseline. We also provide the qualitative results in Fig.~\ref{qualitative}, where our predictions are closer to the ground truth. 

\subsubsection{Ablation Study. }
We conducted ablation experiments to verify the effectiveness of the proposed CFA and UPR. For our baseline, we use ViT-B as the backbone of SAM and remove the prompt encoder. As shown in Table~\ref{tab2}, we are confident to assert that the contribution of each module to the overall performance enhancement is significant, and their combination produces the best overall performance.

 To ensure the reliability of the CFA, we conducted ablation experiments on the introduced cross-branch attention (CA), multi-level features fusion (Fusion), and position embedding replacement (PE). As shown in Table~\ref{tab:table1}, compared with the baseline, our method achieved better performance and improved the mean Dice score by 31.7\% and the mean IoU score by 41.4\%. In addition, we tried to train (1) the Transformer block norm (TN), (2) the neck layer norm (NN), and (3) both to validate the effectiveness of UPR. The results are shown in Table~\ref{tab:table2}. Due to the limitation of computing power, we set the batch size for experiments involving TN to 2, while the others were set to 4. Currently, only training the neck layer norm demonstrated superior performance.

\begin{table}[t]
    \begin{minipage}[]{\textwidth}
    \caption{ Ablation experiments on five public polyp datasets. }\label{tab2}
    \resizebox{\linewidth}{!}{  
    \begin{tabular}{cccccccccccc}
    \toprule
    & \multicolumn{1}{c|}{} & \multicolumn{2}{c}{\textbf{\makebox[2.3cm]{CVC-ClinicDB}}} & \multicolumn{2}{c}{\textbf{\makebox[2cm]{Kvasir}}} & \multicolumn{2}{c}{\textbf{\makebox[2cm]{CVC-ColonDB}}} & \multicolumn{2}{c}{\textbf{\makebox[2cm]{ETIS}}} & \multicolumn{2}{c}{\textbf{\makebox[2cm]{EndoScene}}} \\
    \textbf{CFA} & \multicolumn{1}{c|}{\textbf{UPR}} &Dice & IoU & Dice & IoU & Dice & IoU & Dice & IoU & Dice & IoU \\
    \midrule
    & \multicolumn{1}{c|}{} & 0.537 & 0.388 & 0.652 & 0.469 & 0.397 & 0.268 & 0.387 & 0.265 & 0.525 & 0.368 \\
    \checkmark & \multicolumn{1}{c|}{} & 0.913 & 0.843 & 0.887 & 0.819 & 0.775 & 0.684 & 0.811 & 0.708 & 0.901 & 0.831 \\
    & \multicolumn{1}{c|}{\checkmark} & 0.558 & 0.407 & 0.672 & 0.519 & 0.415 & 0.282 & 0.382 & 0.360 & 0.610 & 0.453 \\
    \checkmark & \multicolumn{1}{c|}{\checkmark} & 0.950 & 0.905 & 0.914 & 0.848 & 0.792 & 0.694 & 0.856 & 0.764 & 0.914 & 0.843 \\
    \bottomrule
    \end{tabular}
    }
    \end{minipage}
  \centering
  \begin{minipage}[t]{0.50\textwidth}
    \caption{Ablation studies on EPA.}
    \centering
    \resizebox{\linewidth}{!}{  
    \begin{tabular}{cccccccc}
    \toprule
    \multicolumn{3}{c}{\textbf{{CFA}}} & \multicolumn{1}{c|}{\textbf{{UPR}}} & \multicolumn{2}{c}{\textbf{\makebox[2.3cm]{CVC-ClinicDB}}} & \multicolumn{2}{c}{\textbf{\makebox[2cm]{Kvasir}}} \\
    \textbf{CA} & \textbf{Fusion} & \textbf{PE} & \multicolumn{1}{c|}{} &  Dice & IoU & Dice & IoU \\
    \midrule
      &   &   & \multicolumn{1}{c|}{\checkmark} & 0.558 & 0.407 & 0.672 & 0.519 \\
    \checkmark &   &   & \multicolumn{1}{c|}{\checkmark} & 0.838 & 0.737 & 0.866 & 0.772 \\
    \checkmark & \checkmark &   & \multicolumn{1}{c|}{\checkmark} & 0.941 & 0.891 & 0.911 & 0.848 \\
    \checkmark & \checkmark & \checkmark & \multicolumn{1}{c|}{\checkmark} & 0.950 & 0.905 & 0.914 & 0.848\\
    \bottomrule
    \end{tabular}
    }
    \label{tab:table1}
  \end{minipage}
  \hfill
  \begin{minipage}[t]{0.46\textwidth}
    \caption{Ablation studies on UPR.}
    \centering
    \resizebox{\linewidth}{!}{ 
    \begin{tabular}{cccccccc}
    \toprule
    \textbf{CFA} & \multicolumn{3}{c|}{\textbf{UPR}} & \multicolumn{2}{c}{\textbf{\makebox[2.3cm]{CVC-ClinicDB}}} & \multicolumn{2}{c}{\textbf{\makebox[2cm]{Kvasir}}} \\
    \textbf{} & \textbf{TN} & \textbf{NN} & \multicolumn{1}{c|}{\textbf{Hint}} &  Dice & IoU & Dice & IoU \\
    \midrule
    \checkmark &   &   & \multicolumn{1}{c|}{} & 0.913 & 0.843 & 0.887 & 0.819 \\
    \checkmark & \checkmark &   & \multicolumn{1}{c|}{} & 0.936 & 0.885 & 0.912 & 0.859 \\
    \checkmark &   & \checkmark & \multicolumn{1}{c|}{} & 0.944 & 0.896 & 0.914 & 0.850 \\
    \checkmark & \checkmark & \checkmark & \multicolumn{1}{c|}{} & 0.936 & 0.881 & 0.875 & 0.814 \\
    \checkmark &   & \checkmark & \multicolumn{1}{c|}{\checkmark} & 0.950 & 0.905 & 0.914 & 0.848 \\
    \bottomrule
    \end{tabular}
    }
    \label{tab:table2}
  \end{minipage}
\end{table}

\section{Conclusion}
We introduce a novel method called ASPS for polyp segmentation, designed to address the limitations of the SAM model in capturing information and bridging the domain gap between endoscopy images. The CFA module incorporates a trainable CNN encoder branch to supplement the frozen ViT encoder, integrating multi-scale and multi-level features. Additionally, the UPR module reduces uncertainty during training by introducing hints and adjusting the normalization layer, promoting the adaptation of the model in the endoscopy field. Through experiments on five common polyp datasets, we verify the effectiveness and superiority of our method. To extend our work, our future direction focuses on investigating more efficient methods using SAM, enabling us to fully harness the foundation model for effective polyp segmentation. 

\begin{credits}
\subsubsection{\ackname} This work is supported by the National Natural Science Foundation of China (No. 62202015, 62322605, 62293543, U21B2048, 62272468, 62306101), Anhui Provincial Key R\&D Programmes (2023s07020001), the University Synergy Innovation Program of Anhui Province (GXXT-2022-052), and Key-Area Research and Development Program of Shaanxi Province under Grant 2023-ZDISF-41.

\subsubsection{\discintname}
We declare that we have no competing interests.
\end{credits}

%
%
%
\bibliographystyle{splncs04}
\bibliography{reference}

\begin{thebibliography}{10}
\providecommand{\url}[1]{\texttt{#1}}
\providecommand{\urlprefix}{URL }
\providecommand{\doi}[1]{https://doi.org/#1}

\bibitem{ba2016layer}
Ba, J.L., Kiros, J.R., Hinton, G.E.: Layer normalization. arXiv preprint
  arXiv:1607.06450  (2016)

\bibitem{bernal2015wm}
Bernal, J., S{\'a}nchez, F.J., Fern{\'a}ndez-Esparrach, G., Gil, D.,
  Rodr{\'\i}guez, C., Vilari{\~n}o, F.: Wm-dova maps for accurate polyp
  highlighting in colonoscopy: Validation vs. saliency maps from physicians.
  Computerized medical imaging and graphics  \textbf{43},  99--111 (2015)

\bibitem{chen2023sam}
Chen, T., Zhu, L., Ding, C., Cao, R., Zhang, S., Wang, Y., Li, Z., Sun, L.,
  Mao, P., Zang, Y.: Sam fails to segment anything?--sam-adapter: Adapting sam
  in underperformed scenes: Camouflage, shadow, and more. arXiv preprint
  arXiv:2304.09148  (2023)

\bibitem{devries2018learning}
DeVries, T., Taylor, G.W.: Learning confidence for out-of-distribution
  detection in neural networks. arXiv preprint arXiv:1802.04865  (2018)

\bibitem{dong2021polyp}
Dong, B., Wang, W., Fan, D.P., Li, J., Fu, H., Shao, L.: Polyp-pvt: Polyp
  segmentation with pyramid vision transformers. arXiv preprint
  arXiv:2108.06932  (2021)

\bibitem{fan2020pranet}
Fan, D.P., Ji, G.P., Zhou, T., Chen, G., Fu, H., Shen, J., Shao, L.: Pranet:
  Parallel reverse attention network for polyp segmentation. In: International
  conference on medical image computing and computer-assisted intervention. pp.
  263--273. Springer (2020)

\bibitem{guo2022segnext}
Guo, M.H., Lu, C.Z., Hou, Q., Liu, Z., Cheng, M.M., Hu, S.M.: Segnext:
  Rethinking convolutional attention design for semantic segmentation. Advances
  in Neural Information Processing Systems  \textbf{35},  1140--1156 (2022)

\bibitem{hoyer2023mic}
Hoyer, L., Dai, D., Wang, H., Van~Gool, L.: Mic: Masked image consistency for
  context-enhanced domain adaptation. In: Proceedings of the IEEE/CVF
  conference on computer vision and pattern recognition. pp. 11721--11732
  (2023)

\bibitem{hu2023efficiently}
Hu, X., Xu, X., Shi, Y.: How to efficiently adapt large segmentation model
  (sam) to medical images. arXiv preprint arXiv:2306.13731  (2023)

\bibitem{jha2020kvasir}
Jha, D., Smedsrud, P.H., Riegler, M.A., Halvorsen, P., de~Lange, T., Johansen,
  D., Johansen, H.D.: Kvasir-seg: A segmented polyp dataset. In: MultiMedia
  Modeling: 26th International Conference, MMM 2020, Daejeon, South Korea,
  January 5--8, 2020, Proceedings, Part II 26. pp. 451--462. Springer (2020)

\bibitem{kim2021uacanet}
Kim, T., Lee, H., Kim, D.: Uacanet: Uncertainty augmented context attention for
  polyp segmentation. In: Proceedings of the 29th ACM International Conference
  on Multimedia. pp. 2167--2175 (2021)

\bibitem{kirillov2023segment}
Kirillov, A., Mintun, E., Ravi, N., Mao, H., Rolland, C., Gustafson, L., Xiao,
  T., Whitehead, S., Berg, A.C., Lo, W.Y., et~al.: Segment anything. arXiv
  preprint arXiv:2304.02643  (2023)

\bibitem{li2016revisiting}
Li, Y., Wang, N., Shi, J., Liu, J., Hou, X.: Revisiting batch normalization for
  practical domain adaptation. arXiv preprint arXiv:1603.04779  (2016)

\bibitem{lin2023samus}
Lin, X., Xiang, Y., Zhang, L., Yang, X., Yan, Z., Yu, L.: Samus: Adapting
  segment anything model for clinically-friendly and generalizable ultrasound
  image segmentation. arXiv preprint arXiv:2309.06824  (2023)

\bibitem{ma2024segment}
Ma, J., He, Y., Li, F., Han, L., You, C., Wang, B.: Segment anything in medical
  images. Nature Communications  \textbf{15}(1), ~654 (2024)

\bibitem{nguyen2023out}
Nguyen, D.M.H., Pham, T.N., Diep, N.T., Phan, N.Q., Pham, Q., Tong, V., Nguyen,
  B.T., Le, N.H., Ho, N., Xie, P., et~al.: On the out of distribution
  robustness of foundation models in medical image segmentation. arXiv preprint
  arXiv:2311.11096  (2023)

\bibitem{pan2022fast}
Pan, Z., Cai, J., Zhuang, B.: Fast vision transformers with hilo attention.
  Advances in Neural Information Processing Systems  \textbf{35},  14541--14554
  (2022)

\bibitem{ronneberger2015u}
Ronneberger, O., Fischer, P., Brox, T.: U-net: Convolutional networks for
  biomedical image segmentation. In: Medical Image Computing and
  Computer-Assisted Intervention--MICCAI 2015: 18th International Conference,
  Munich, Germany, October 5-9, 2015, Proceedings, Part III 18. pp. 234--241.
  Springer (2015)

\bibitem{silva2014toward}
Silva, J., Histace, A., Romain, O., Dray, X., Granado, B.: Toward embedded
  detection of polyps in wce images for early diagnosis of colorectal cancer.
  International journal of computer assisted radiology and surgery  \textbf{9},
   283--293 (2014)

\bibitem{tajbakhsh2015automated}
Tajbakhsh, N., Gurudu, S.R., Liang, J.: Automated polyp detection in
  colonoscopy videos using shape and context information. IEEE transactions on
  medical imaging  \textbf{35}(2),  630--644 (2015)

\bibitem{vazquez2017benchmark}
V{\'a}zquez, D., Bernal, J., S{\'a}nchez, F.J., Fern{\'a}ndez-Esparrach, G.,
  L{\'o}pez, A.M., Romero, A., Drozdzal, M., Courville, A., et~al.: A benchmark
  for endoluminal scene segmentation of colonoscopy images. Journal of
  healthcare engineering  \textbf{2017} (2017)

\bibitem{wang2022stepwise}
Wang, J., Huang, Q., Tang, F., Meng, J., Su, J., Song, S.: Stepwise feature
  fusion: Local guides global. In: International Conference on Medical Image
  Computing and Computer-Assisted Intervention. pp. 110--120. Springer (2022)

\bibitem{wang2023foundation}
Wang, Z., Liu, C., Zhang, S., Dou, Q.: Foundation model for endoscopy video
  analysis via large-scale self-supervised pre-train. In: International
  Conference on Medical Image Computing and Computer-Assisted Intervention. pp.
  101--111. Springer (2023)

\bibitem{wei2021shallow}
Wei, J., Hu, Y., Zhang, R., Li, Z., Zhou, S.K., Cui, S.: Shallow attention
  network for polyp segmentation. In: Medical Image Computing and Computer
  Assisted Intervention--MICCAI 2021: 24th International Conference,
  Strasbourg, France, September 27--October 1, 2021, Proceedings, Part I 24.
  pp. 699--708. Springer (2021)

\bibitem{xiong2023efficientsam}
Xiong, Y., Varadarajan, B., Wu, L., Xiang, X., Xiao, F., Zhu, C., Dai, X.,
  Wang, D., Sun, F., Iandola, F., et~al.: Efficientsam: Leveraged masked image
  pretraining for efficient segment anything. arXiv preprint arXiv:2312.00863
  (2023)

\bibitem{yang2021context}
Yang, J., An, W., Yan, C., Zhao, P., Huang, J.: Context-aware domain adaptation
  in semantic segmentation. In: Proceedings of the IEEE/CVF Winter Conference
  on Applications of Computer Vision. pp. 514--524 (2021)

\bibitem{yin2022duplex}
Yin, Z., Liang, K., Ma, Z., Guo, J.: Duplex contextual relation network for
  polyp segmentation. In: 2022 IEEE 19th International Symposium on Biomedical
  Imaging (ISBI). pp.~1--5. IEEE (2022)

\bibitem{yue2023surgicalsam}
Yue, W., Zhang, J., Hu, K., Xia, Y., Luo, J., Wang, Z.: Surgicalsam: Efficient
  class promptable surgical instrument segmentation. arXiv preprint
  arXiv:2308.08746  (2023)

\bibitem{zhang2023sam}
Zhang, J., Ma, K., Kapse, S., Saltz, J., Vakalopoulou, M., Prasanna, P.,
  Samaras, D.: Sam-path: A segment anything model for semantic segmentation in
  digital pathology. In: International Conference on Medical Image Computing
  and Computer-Assisted Intervention. pp. 161--170. Springer (2023)

\bibitem{zhang2023customized}
Zhang, K., Liu, D.: Customized segment anything model for medical image
  segmentation. arXiv preprint arXiv:2304.13785  (2023)

\bibitem{zhang2022lesion}
Zhang, R., Lai, P., Wan, X., Fan, D.J., Gao, F., Wu, X.J., Li, G.: Lesion-aware
  dynamic kernel for polyp segmentation. In: International Conference on
  Medical Image Computing and Computer-Assisted Intervention. pp. 99--109.
  Springer (2022)

\bibitem{zhao2023tuning}
Zhao, B., Tu, H., Wei, C., Mei, J., Xie, C.: Tuning layernorm in attention:
  Towards efficient multi-modal llm finetuning. arXiv preprint arXiv:2312.11420
   (2023)

\bibitem{zhao2021automatic}
Zhao, X., Zhang, L., Lu, H.: Automatic polyp segmentation via multi-scale
  subtraction network. In: Medical Image Computing and Computer Assisted
  Intervention--MICCAI 2021: 24th International Conference, Strasbourg, France,
  September 27--October 1, 2021, Proceedings, Part I 24. pp. 120--130. Springer
  (2021)

\bibitem{zhou2023can}
Zhou, T., Zhang, Y., Zhou, Y., Wu, Y., Gong, C.: Can sam segment polyps? arXiv
  preprint arXiv:2304.07583  (2023)

\bibitem{zhou2023cross}
Zhou, T., Zhou, Y., He, K., Gong, C., Yang, J., Fu, H., Shen, D.: Cross-level
  feature aggregation network for polyp segmentation. Pattern Recognition
  \textbf{140},  109555 (2023)

\end{thebibliography}

\end{document}